\documentclass[a4paper,11pt]{article}

\usepackage{caption}
\usepackage{subcaption}

\usepackage{rotating}

\usepackage{jinstpub} 
\usepackage[export]{adjustbox}

\usepackage{siunitx}
\DeclareSIUnit\bar{bar}

\title{First operation of an ACHINOS-equipped Spherical Proportional Counter with individual anode read-out}

\author[a]{D.~Herd,}
\author[a, b]{I.~Katsioulas,}
\author[a, 1]{P.~Knights,\note{Corresponding author.}}
\author[a]{I.~Manthos,} 
\author[a]{J.~Matthews,}
\author[a, c]{L.~Millins,}
\author[a]{T.~Neep,}
\author[a,d]{K.~Nikolopoulos,}
\author[a]{G.~Rogers} 

\affiliation[a]{School of Physics and Astronomy, University of Birmingham, Birmingham, B15 2TT, United Kingdom}
\affiliation[b]{European Spallation Source ESS ERIC (ESS), Lund, SE-221 00 , Sweden}
\affiliation[c]{Particle Physics Department, STFC Rutherford Appleton Laboratory, Didcot, OX11 0QX, UK}
\affiliation[d]{Institute for Experimental Physics, University of Hamburg, Hamburg, 22761, Germany}
\emailAdd{p.r.knights@bham.ac.uk}

\abstract{The multi-anode sensor ACHINOS revolutionised the capabilities of the spherical proportional counter by enabling large-size, high-pressure, operation and TPC-like reconstruction capabilities through individual anode read-out. First measurements with an individually read out ACHINOS are performed, which enables improved calibration and response homogenisation. Experimental results demonstrating the improvement in energy resolution brought by the individual anode calibration are presented. These are complemented by detailed simulation studies on the effect of sensor design and manufacturing imperfections, and how they may be corrected both in hardware and analysis.}


\begin{document}
\maketitle
\flushbottom

\section{Introduction}
\label{sec:intro}
\begin{figure}[htbp]
\centering
\includegraphics[width=0.7\textwidth]{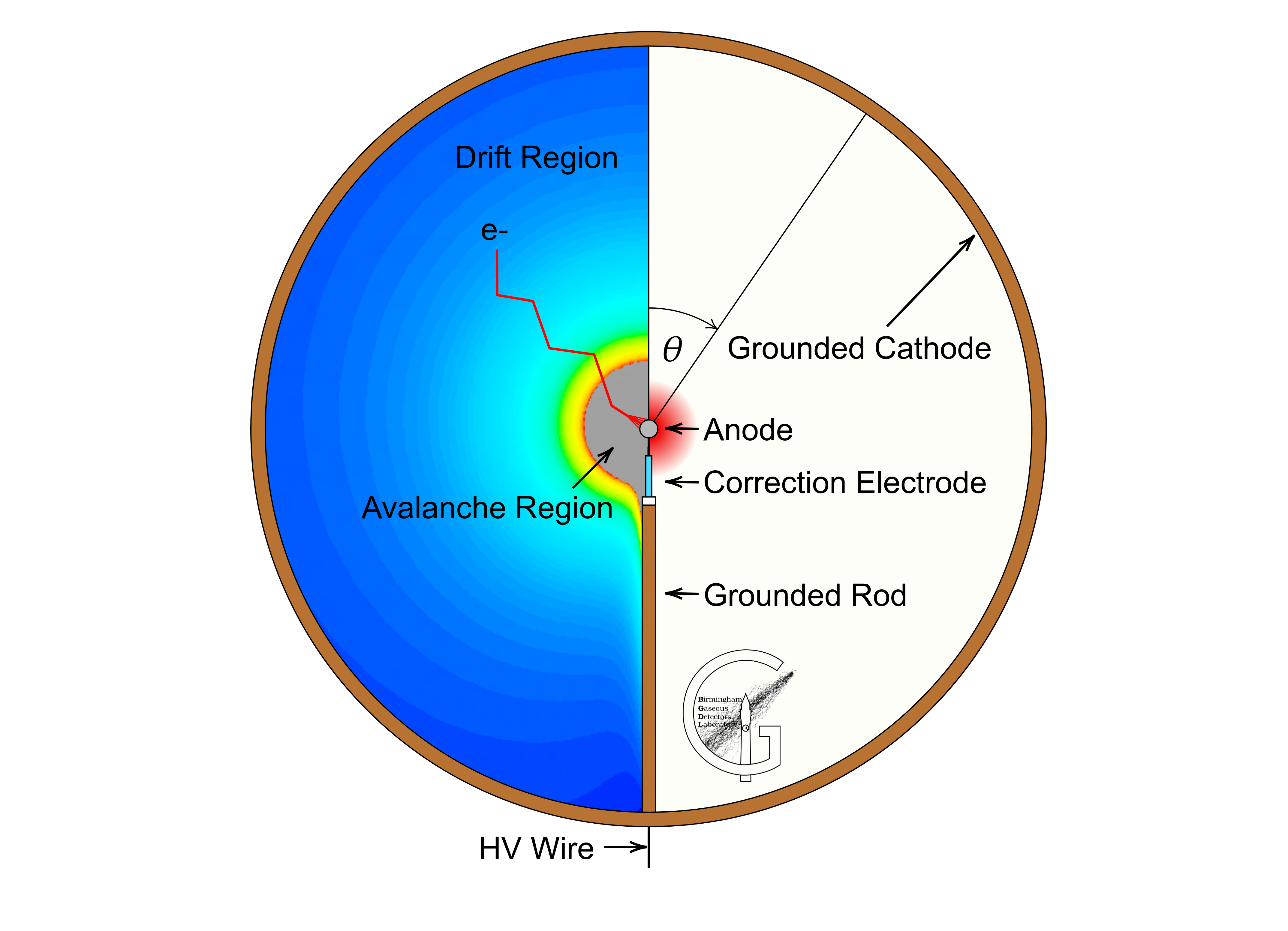}
\caption{\label{fig:sphericalproportionacounter} Schematic of the spherical proportional counter showing the grounded cathode and the central electrode and its support structure.}
\end{figure}

The spherical proportional counter~\cite{Giomataris2008-mx,Katsioulas:2018pyhSUB} is a gaseous detector with wide-ranging applications from neutron spectroscopy~\cite{Giomataris:2022bvz} to direct dark matter searches~\cite{Arnaud2018-nr,NEWS-G:2021vfh}. The detector in its simplest form is shown in Figure~\ref{fig:sphericalproportionacounter}. It  comprises an $\mathcal{O}(\si{\meter})$ in radius grounded spherical shell, which acts as the cathode, and an $\mathcal{O}(\si{\milli\meter})$ in radius spherical anode at its centre, to which a high voltage $V$ is applied and from which the signal is read out. Neglecting the structures that hold the anode in place, the electric field $\vec{E}(r)$ as a function of radius $r$ is given by
\begin{equation}
    \vec{E}(r) = \frac{V}{r^2} \frac{r_{a} r_{c}}{r_{c}-r_{a}} \hat{r}\approx \frac{V}{r^2} r_{a} \hat{r}\,,
    \label{eq:field}
\end{equation}
where $r_{c}$ and $r_{a}$ are the shell and anode radii respectively. 
The $1/r^{2}$ dependence means that the electric field becomes sufficiently large near the anode to cause drifting electrons to avalanche, providing signal amplification.

\begin{figure}[htbp]
\centering
     \begin{subfigure}[b]{0.35\textwidth}
         \centering
         \includegraphics[width=\textwidth]{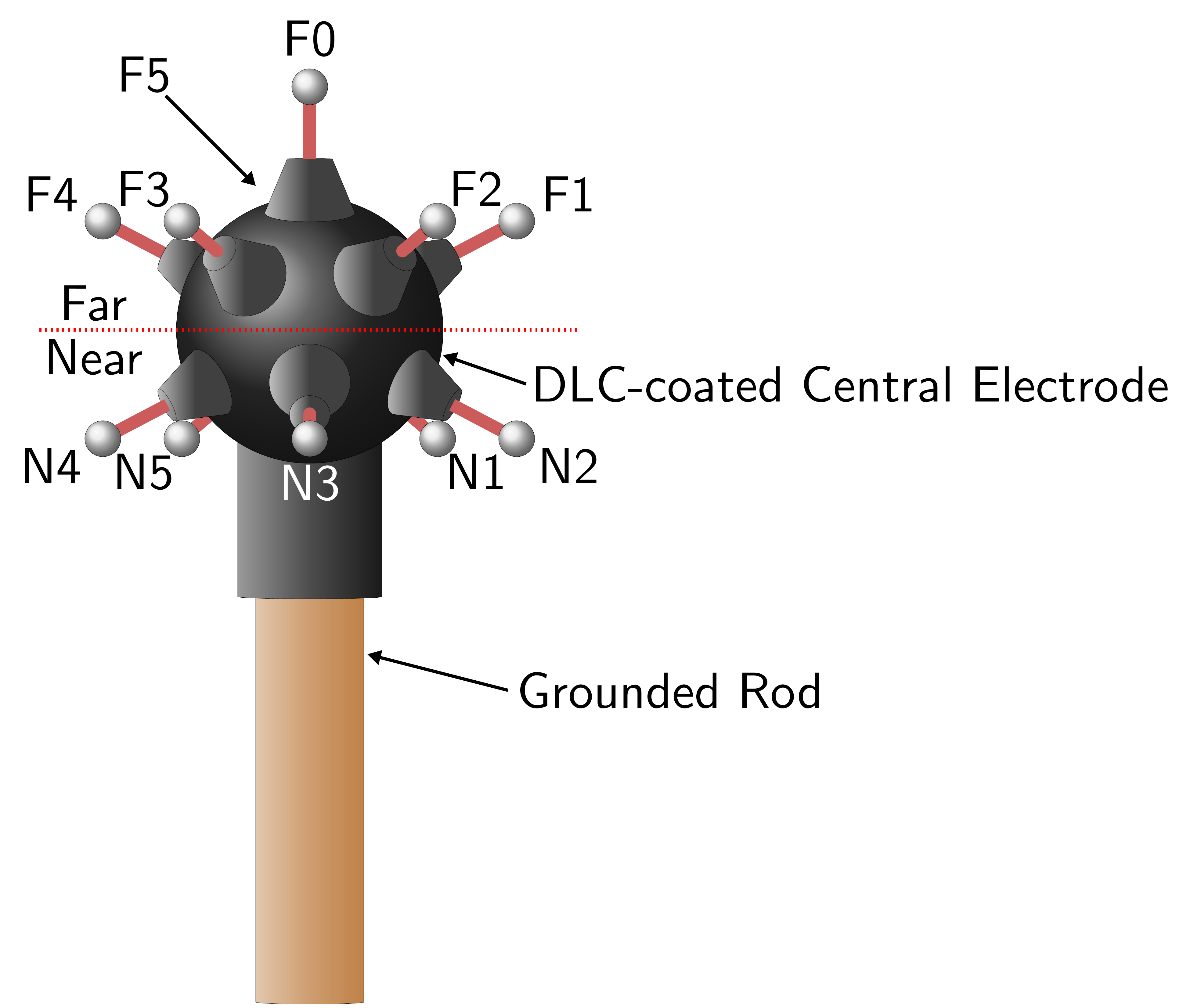}
         \caption{\label{subfig:achinosNearFar}}
     \end{subfigure}
     \begin{subfigure}[b]{0.22\textwidth}
         \centering
         \includegraphics[width=\textwidth]{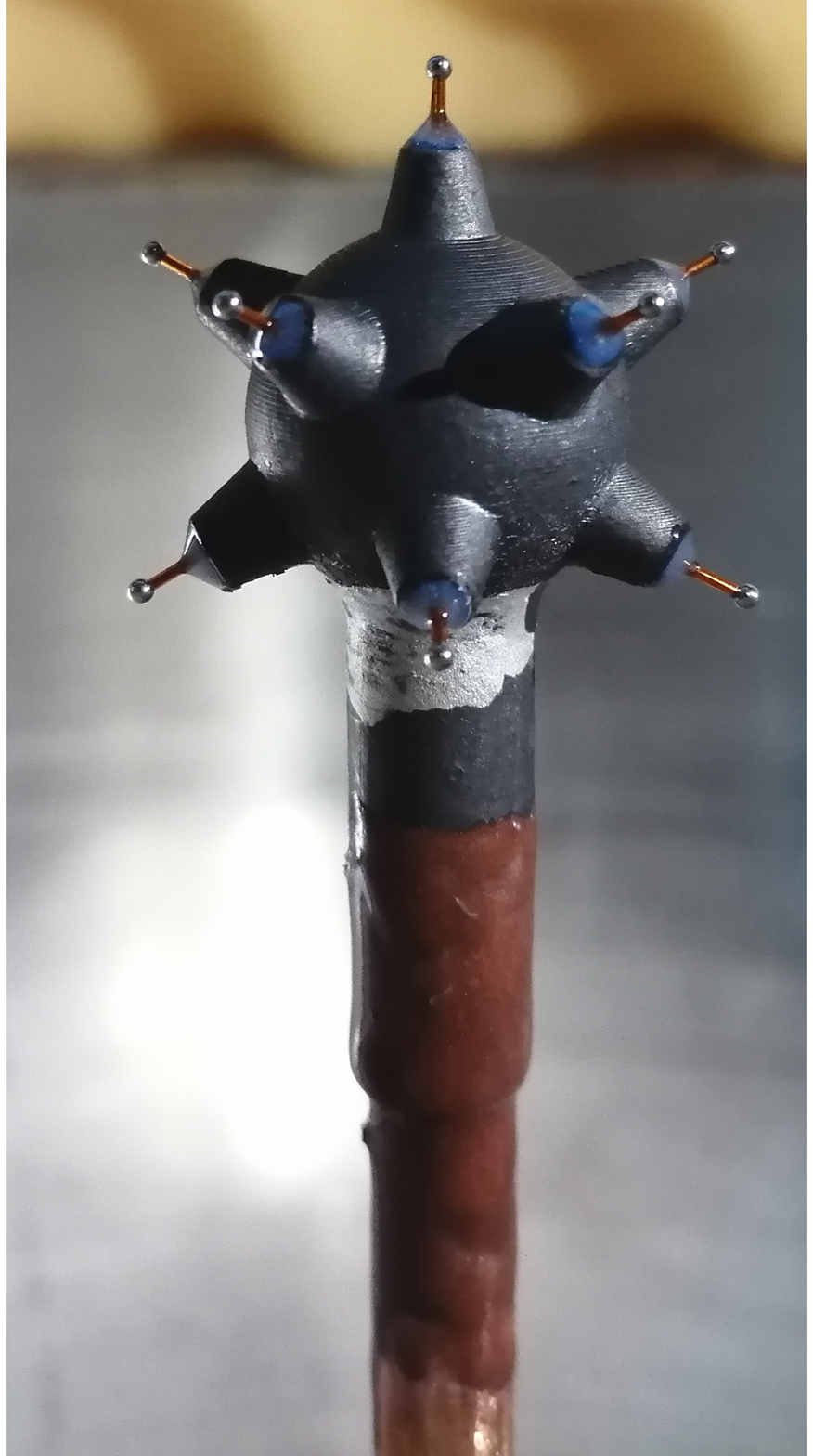}
         \caption{\label{subfig:achinosPhoto}}
     \end{subfigure}
     \begin{subfigure}[b]{0.33\textwidth}
         \centering
         \includegraphics[width=\textwidth]{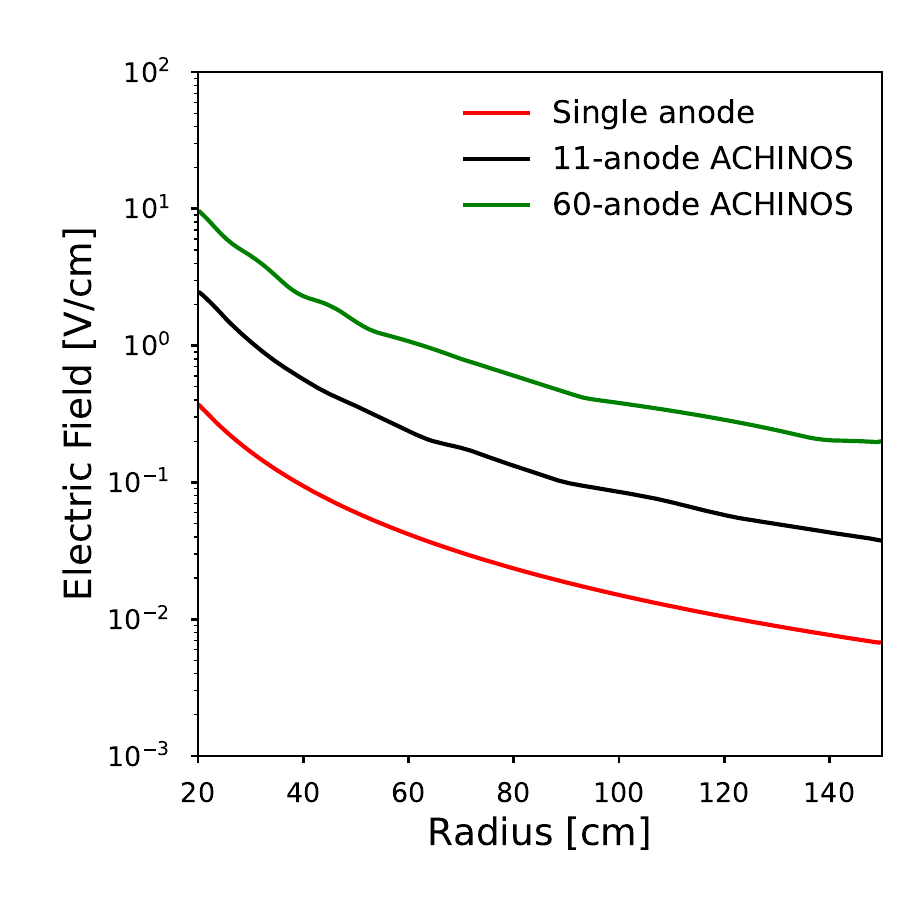}
         \caption{\label{subfig:achinosFieldComparison}}
     \end{subfigure}
\caption{\label{fig:achinosSchematics} (\subref{subfig:achinosNearFar}) Schematic of an 11-anode ACHINOS. The anodes are located at the vertices of an icosahedron. Current implementations of ACHINOS use either single-channel read-out, where the signal is read from all anodes together, or two-channel read-out, with the anodes being grouped into the five `Near' and six `Far', based on their relative distance from the rod. The individual anodes are labelled for reference. (\subref{subfig:achinosPhoto}) Example of an 11-anode ACHINOS. (\subref{subfig:achinosFieldComparison}) Comparison of the electric field magnitude as a function of radius in the detector for a single anode, 11-anode, and a 60-anode sensor. All sensors have $r_{a}=1\;\si{\milli\meter}$, and the 11-anode (60-anode) ACHINOS have $r_{A}=1.4\;\si{\centi\meter}$ ($r_{A}=15\;\si{\centi\meter}$).}
\end{figure}

Many of the spherical proportional counter's applications require increasingly large detector volume, higher-pressure operation, or both. For example, the NEWS-G collaboration employing the spherical proportional counter for direct dark matter searches would benefit from increased target mass in the detector. In such cases, the single anode can present a limitation: the drift field, which collects the electrons from the detector volume, and the avalanche field are coupled with both being predominantly determined by the anode radius and voltage, as demonstrated in Eq.~\ref{eq:field}. To maintain efficient electron collection in larger volume detectors or in higher pressures, a commensurate increase in the anode radius and its voltage would be required, as demonstrated by Equation~\ref{eq:field}. However, this can negatively impact the stability of the detector, increasing the probability of electric discharges. 
To overcome this limitation, the multi-anode read-out structure, ACHINOS, has been developed~\cite{Giganon2017-jinst,Giomataris:2020rna, Katsioulas:2022cqe}. Shown in Figure~\ref{fig:achinosSchematics}, ACHINOS comprises several $\mathcal{O}(\si{\milli\meter})$ in radius anodes situated at the same distance $r_{\textrm{A}}$ from the centre of the detector. In this way, the drift and avalanche fields are decoupled. To leading order, the avalanche field is determined by the radius and voltage of the individual anode to which the electron arrives, whereas the drift field that the electron experiences in the detector volume is determined by the collective field of all the anodes. Thus, by tuning $r_{\textrm{A}}$ and the number of anodes the drift field can be chosen for a given anode radius and voltage. This is demonstrated in Figure~\ref{subfig:achinosFieldComparison}.

In previous implementations, the read-out of ACHINOS has been constrained to a single- or two-channel configuration due to limitations in the instrumentation and electronics chain, suppressing some of the benefits that ACHINOS can offer. In this article the implementation of individual ACHINOS anode read-out is discussed. In Section~\ref{sec:achinos}, the operation, understanding, and limitations of previous implementations are summarised with the benefits from individual anode read-out. In Section~\ref{sec:setup}, the experimental set-up is described, and in Section~\ref{sec:Experimentalresults} experimental individual anode calibration results are presented. In section~\ref{sec:simulation}, the sources of response non-uniformity are described, using simulations to study the effect of construction and assembly imperfections on avalanche gain. The simulations are subsequently used to show how these effects may be accounted for by individually biasing the anodes.

\section{Existing ACHINOS implementations}
\label{sec:achinos}
Most ACHINOS used in applications of the spherical proportional counter to date have employed 11-anodes positioned at the vertices of a virtual icosahedron, the twelfth vertex being occupied by the grounded support rod, as shown in Figure~\ref{fig:achinosSchematics}. The anodes are placed around a central electrode that shapes the electric field and provides mechanical support. The central electrode comprises a 3D-printed structure coated in a resistive material, typically diamond-like carbon (DLC)~\cite{Giomataris:2020rna}. In the following, the current capabilities of  ACHINOS and how the presented work builds upon these are discussed. 

\subsection{Electric field homogeneity}
The anodes are typically grouped together into either one or two read-out channels. In the two-channel case, the anodes are grouped as shown in Figure~\ref{fig:achinosSchematics} - namely, into the `Near' and `Far' channels, describing the 5 anodes nearest the rod, and the 6 anodes further away, respectively. It has been shown that due to the proximity of the rod, the Near anodes are expected to have a higher electric field at their surface, and, thus, exhibit a higher gain~\cite{Katsioulas:2022cqe}. Furthermore, the difference in the electric field is expected to affect the drift and diffusion of electrons, influencing the pulse-shape parameters used in data analysis. To account for this, the two channels typically have their high-voltages supplied separately, allowing different voltages to be applied and compensate for the gain difference. Furthermore, by reading them out separately, the difference in pulse-shape parameters can be recorded and accounted for in the analysis. ACHINOS operated in this way have been employed in several measurements with spherical proportional counter, including direct dark matter searches performed by the NEWS-G collaboration. 

The difference in the electric field between the Near and Far anodes causes the main gain difference between anodes, however, it is not the only source of such difference. It has been noted in Ref.~\cite{Giomataris:2020rna} that the difference in signal amplitude observed for one of the measurement locations was likely due to a difference in the gain of a specific anode. In Ref.~\cite{Giomataris:2022bvz}, a similar effect was observed and was confirmed through simulations as being due to one of the anodes having a different gain. If, however, the signal from each anode was read-out separately, these gain differences could be accounted for by reading out and biasing each anode individually, allowing for individual anode calibration. 

\subsection{Event localisation}
For applications of the detector where event localisation is important, for example, detector fiducialisation in rare-event searches, pulse-shape analysis can be used to provide some information. In the case of the single-anode read-out, the radial coordinate of a `point-like' interaction, or whether an interaction is point-like or an extended track of energy deposition, is encoded in the pulse rise-time, defined as the time between the $10\%$ and $90\%$ of the signal amplitude. For interactions occurring at larger radii in the detector, primary electrons undergo more diffusion resulting in a greater spread in their arrival time at the anode and an increased risetime. Similarly, extended tracks of ionisation in the detector result in a higher risetime. 

In addition to this radial information, ACHINOS provides the potential to localise the event in the detector. In the two-channel read-out of ACHINOS, it has been shown that the amplitude asymmetry between the Near and Far signals can provide information as to the polar angle of the interaction~\cite{Katsioulas:2022cqe}. This could be further improved by individually reading out each anode of the ACHINOS, using both the individual pulse-shape characteristics and timing information, which would provide radial, polar, and azimuthal information for interactions. This would significantly enhance the spherical proportional counter's fiducialisation capability. Furthermore, this opens up the possibility of track reconstruction for particles leaving an extended track of ionisation in the detector.
This would provide another handle for background rejection in rare event search applications. Additionally, it could provide useful particle identification capabilities, such as distinguishing protons from $\alpha$-particles in spherical proportional counters used for fast neutron spectroscopy~\cite{Giomataris:2022bvz}.

\subsection{Induced signal}
\label{sec:inducedSignal}
\begin{figure}[htbp]
\centering
\includegraphics[width=1\textwidth]{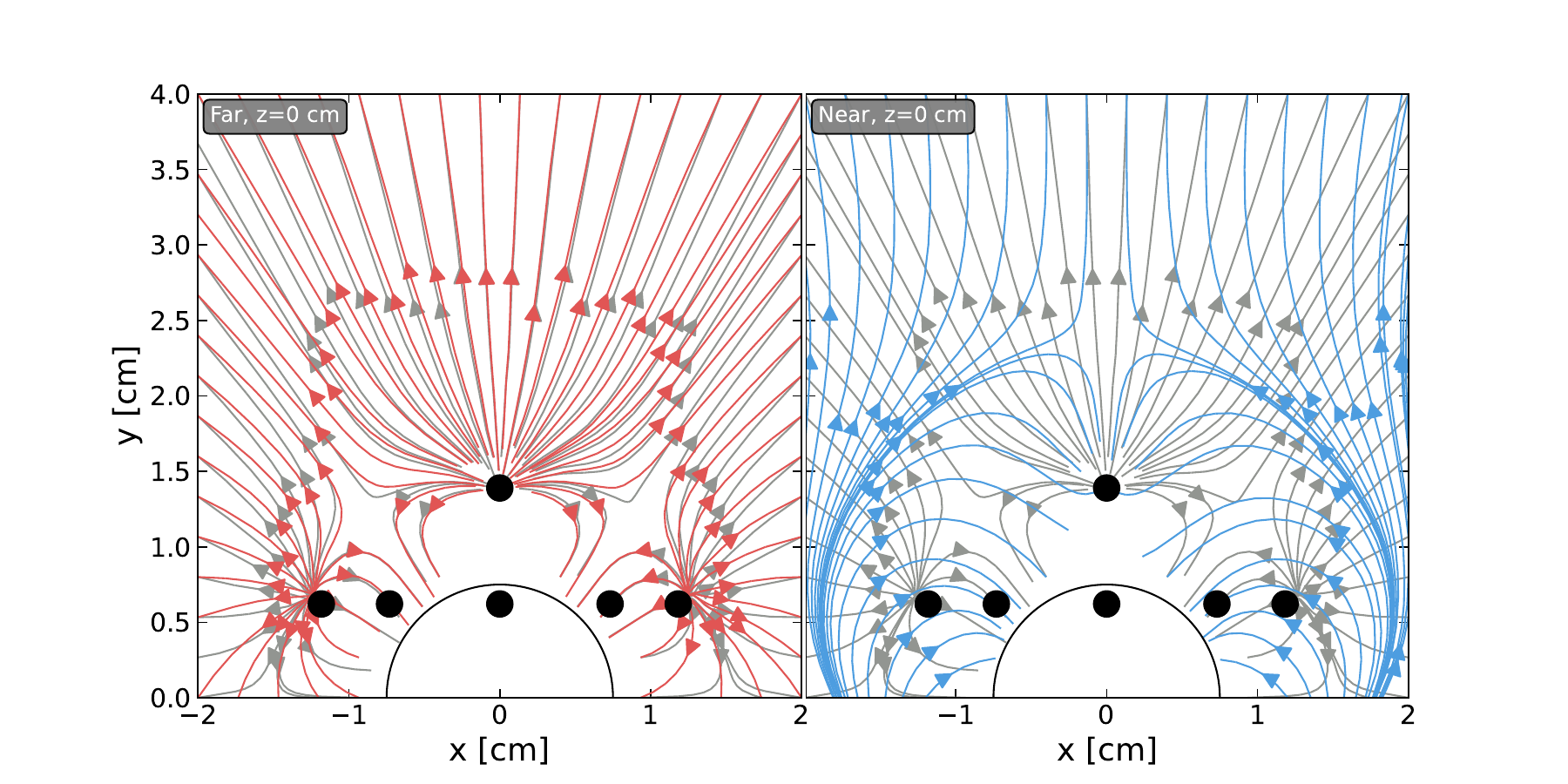}
\caption{\label{fig:achinosweightingfields} The electric (grey lines) and weighting fields of the Far anodes (red) and Near anodes (blue) in the vicinity of the far anodes (black circles). At the surface of the anodes, the two weighting fields are approximately anti-parallel.}
\end{figure}

For a charge $q$ moving in the detector, the instantaneous current $i$ induced on an electrode can be computed using the Shockley-Ramo theorem, 
\begin{equation}
    i = -q \frac{\vec{E}_{W}(\vec{r})\cdot \vec{v}}{V_{W}}\,,
\end{equation}
where $\vec{v}$ is the instantaneous particle velocity, $\vec{E}_{W}$ is the weighting field of the electrode of interest as a function of the particle position $\vec{r}$, and $V_{W}$ is the respective weighting potential.
For example, for electrons arriving at the surface of the Far anodes, it can be seen that the Far and Near weighting fields are approximately anti-parallel, as shown in Figure~\ref{fig:achinosweightingfields}.
Thus, ionisation electrons from a particle interaction in the gas volume that create an avalanche at the Far anodes induce a signal of opposite polarity on the Near and Far anodes respectively.
This can be extended to the case where each anode is read out individually, and so there are 11 weighting fields used in the signal calculation. In the specific case where all ionisation electrons arrive to a single anode, all other anodes will exhibit an opposite polarity signal. 

\section{Experimental set-up and electronics chain}
\label{sec:setup}
\begin{figure}[htbp]
\centering
\includegraphics[width=1\textwidth]{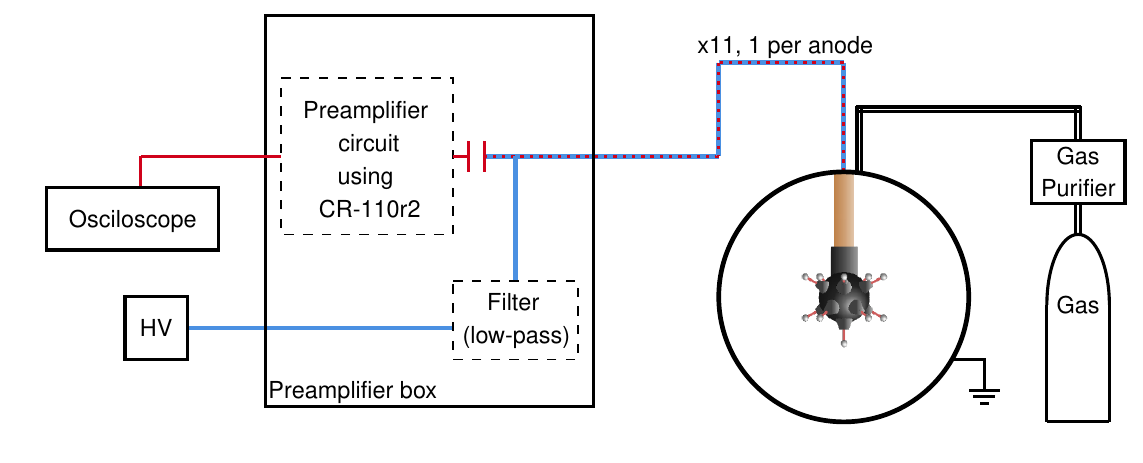}
\caption{\label{fig:experimentalSetup} Schematic of the experimental set-up. The signal from each anode is passed to a preamplifier box, which supplies the high-voltage to the anode and decouples it from the preamplifier circuit. The amplified signal of each anode is passed to an oscilloscope for read-out.}
\end{figure}

The individual anode read-out of the ACHINOS was studied experimentally with the set-up shown in Figure~\ref{fig:experimentalSetup}. An ACHINOS with eleven, $0.5\;\si{\milli\meter}$ in radius, anodes was installed in a $15\;\si{\centi\meter}$ in radius spherical proportional counter. The signal cable of each anode was connected to a high-voltage feed-through on the detector's vacuum system and the anode signal was read out using a purpose-made preamplifier board with a CREMAT CR110-r2 preamplifier card. 
The preamplifier signals were recorded using oscilloscopes. The anodes were each supplied with high voltage independently. This specific multi-anode sensor was selected for the first exploration of the individual anode read-out because it was known to have significantly worse response uniformity performance than the typical sensor.

To account for possible variations between the eleven preamplifier circuits used, each was individually calibrated using a square-wave pulse injected to the input via a $1\;\si{\pico\farad}$ capacitor. The difference in gain between the preamplifiers, which was not greater than $5\%$, was included in subsequent analysis.

A $^{210}$Po source was placed inside the detector onto the inner surface of the cathode. $^{210}$Po decays via $\alpha$-decay, primarily releasing a $5.3\;\si{\mega\eV}$ $\alpha$ particle. The position of the source on the cathode could be adjusted without opening the detector vessel.

Prior to filling the detector with gas, it was evacuated to a pressure of $\num{4E-6}\;\si{\milli\bar}$ using a vacuum pump. Gas was then introduced into the detector through a purpose-built low-radon emanation purifier~\cite{Altenmuller:2021fly} to remove impurities such as oxygen and water vapour.

\section{Experimental results}
\label{sec:Experimentalresults}
The detector was filled with $500\;\si{\milli\bar}$ of Ar:CH$_{4}$ ($98\%$:$2\%$), and the anodes had a voltage of $800\;\si{\volt}$ applied to them, which was estimated to correspond to a gas gain of approximately 30.

\begin{figure}[htbp]
\centering
\includegraphics[width=0.4\textwidth]{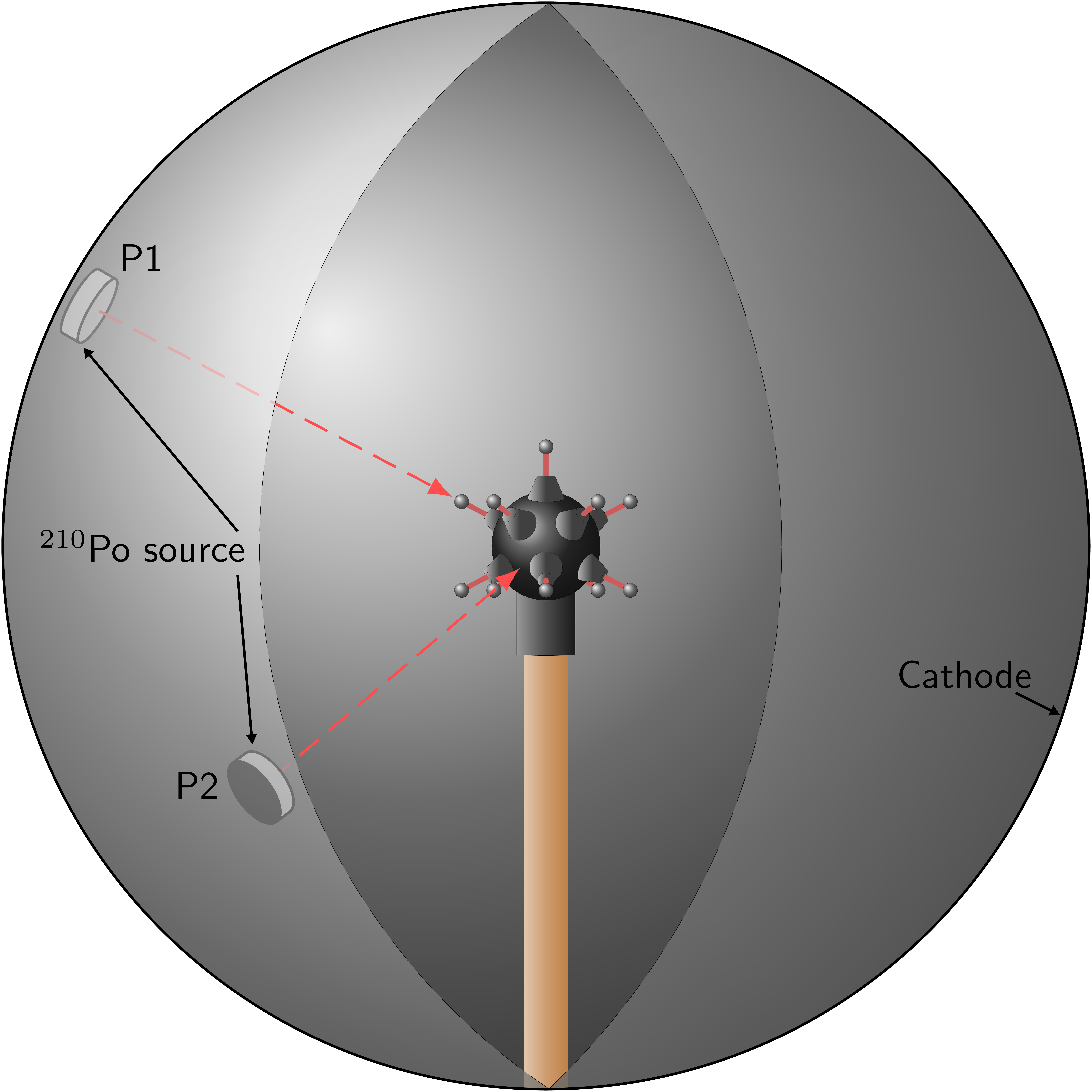}
\caption{\label{fig:achinosCalibrationDiagram} Schematic of the experimental set-up, showing example $^{210}$Po source positions on the internal cathode surface. `P1' shows an example position when the source was adjusted to sit on the cathode directly in line with an anode, whereas `P2' is an example position aligned between two anodes.}
\end{figure}

The anodes were first individually calibrated. 
The $^{210}$Po source was directed at each anode, as shown in Figure~\ref{fig:achinosCalibrationDiagram}, in order to calibrate the response of each anode separately. 
To ensure that the calibration was representative of the signal when the full ionisation charge is collected on the anode at which the source is pointed, a selection was placed on the signal amplitudes such that only events where all other anodes had opposite polarity signals were used in the calibration, as explained in Section~\ref{sec:inducedSignal}.

\begin{figure}[htbp]
\includegraphics[width=0.99\textwidth]{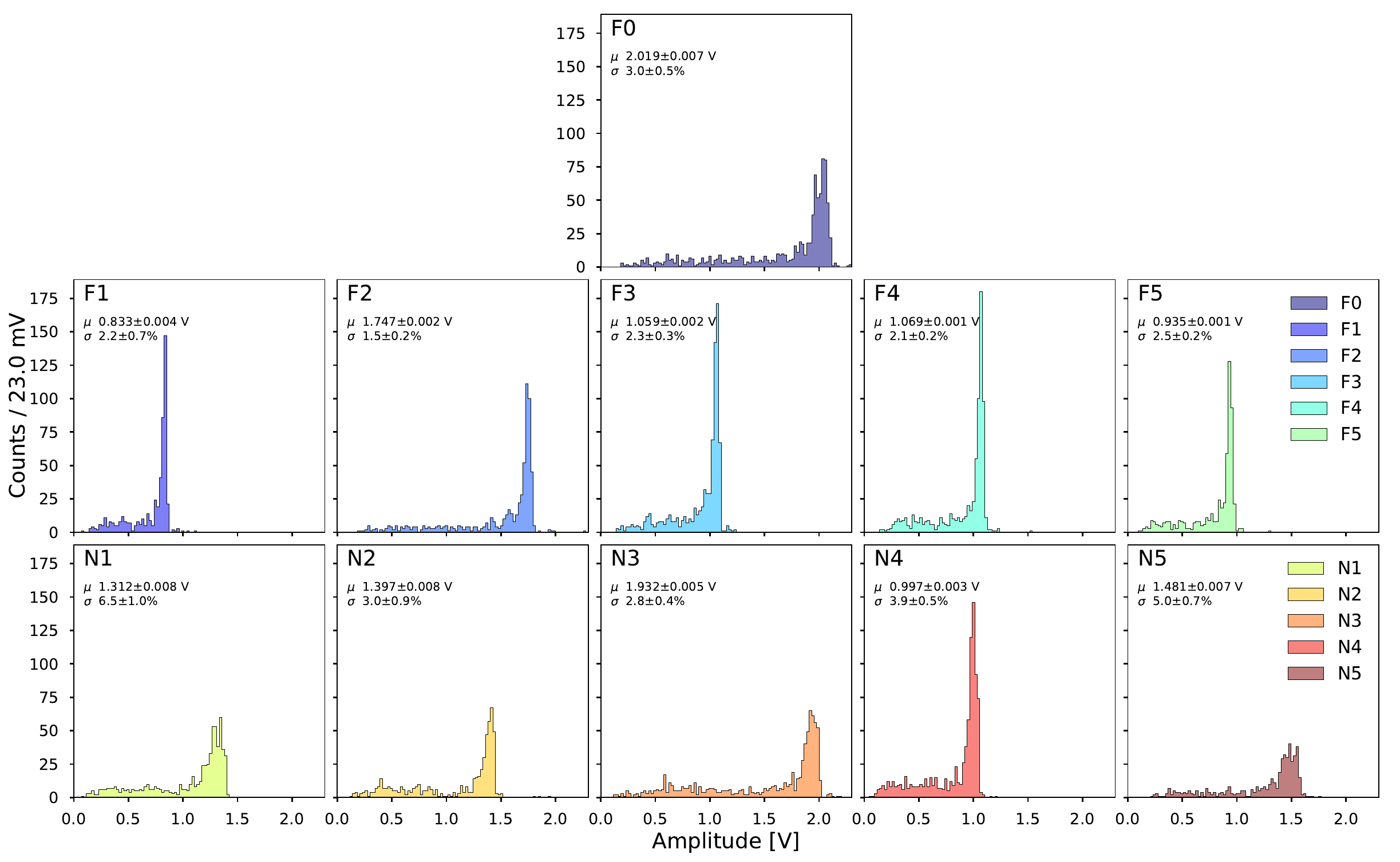}
\caption{\label{fig:allAnodescalibComparison} The amplitude distributions used for the calibration. Each panel shows the  response of a given anode, indicated in the top-left of the panel, when the source is directly aligned with it. Data are selected to remove events where charge is shared between multiple anodes, as described in the text. The distributions are fit with a Gaussian function to obtain an estimate for the mean and local energy resolution, and used as the calibration for each anode.}
\end{figure}

The pulse amplitude distributions recorded from each anode are shown in Figure~\ref{fig:allAnodescalibComparison}.  The observed peak in amplitude was fit with a Gaussian function to estimate the anode's gain and local energy resolution.  

\begin{figure}[htbp]
\centering
 \begin{subfigure}[t]{0.495\textwidth}
         \centering
         \includegraphics[width=\textwidth]{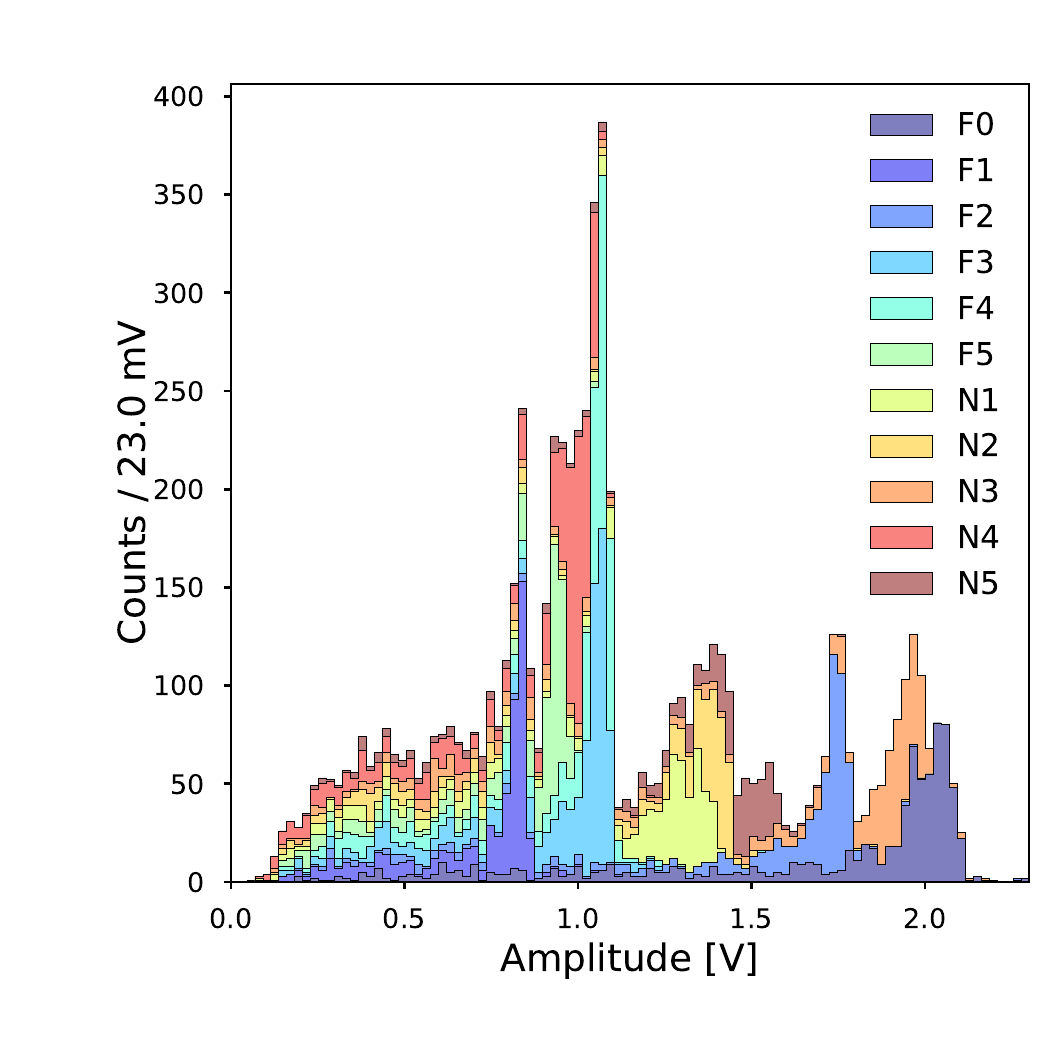}
         \caption{\label{subfig:allAnodesuncalibAmps}}
     \end{subfigure}
     \begin{subfigure}[t]{0.495\textwidth}
         \centering
         \includegraphics[width=\textwidth]{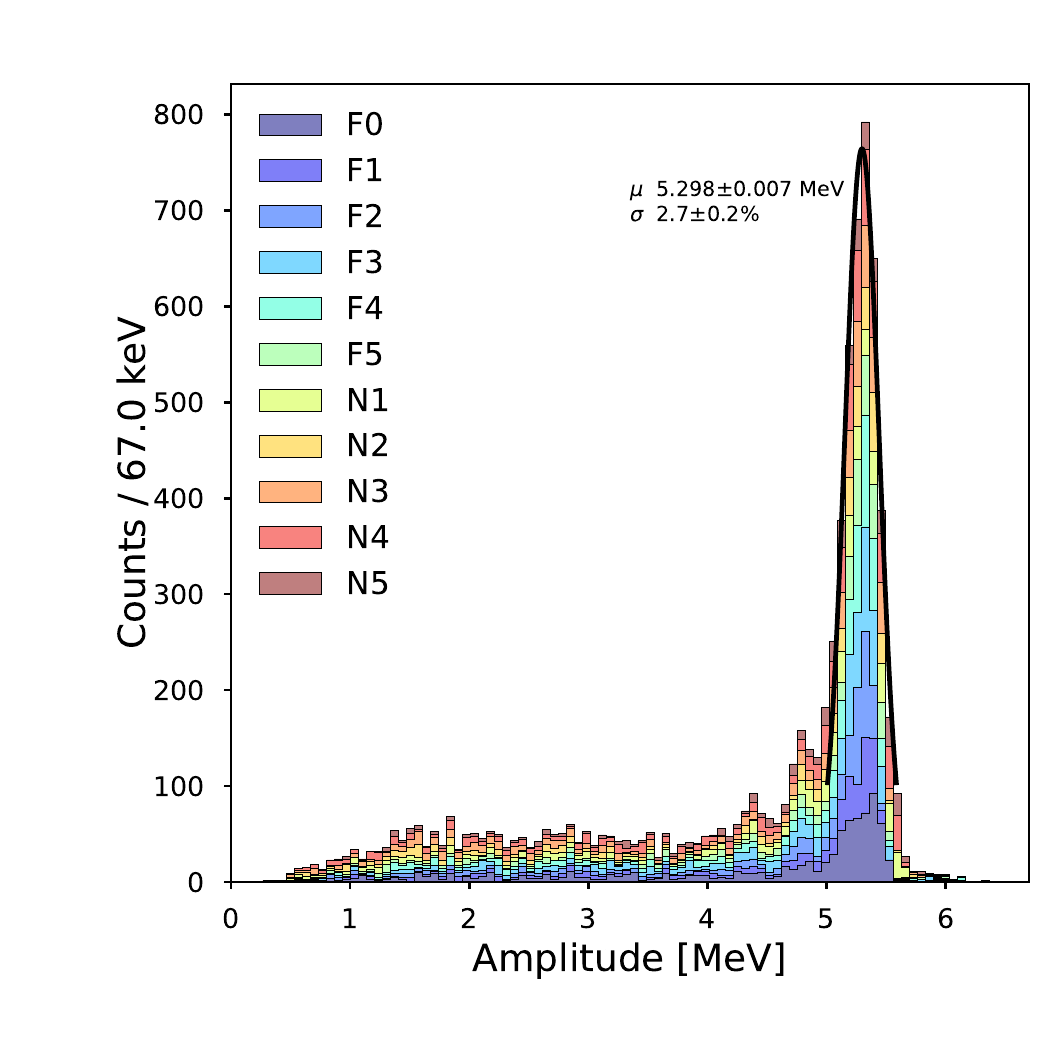}
         \caption{\label{subfig:allAnodescalibAmps}}
     \end{subfigure}
\caption{\label{fig:stackedAmpes} Stacked amplitude distributions from the individual calibration plots shown in Figure~\ref{fig:allAnodescalibComparison} when the individual distributions are  (\subref{subfig:allAnodesuncalibAmps}) uncalibrated and (\subref{subfig:allAnodescalibAmps}) calibrated. The calibrated case has been fit with a Gaussian function (black line) to  estimate the energy resolution.} 
\end{figure}

To estimate the difference in energy resolution when individual anode calibration is employed, the amplitude distributions in Figure~\ref{fig:allAnodescalibComparison} are stacked in Figure~\ref{fig:stackedAmpes} with and without the calibration applied. A single Gaussian does not describe the uncalibrated data, however, the energy resolution after calibration was estimated by fitting the overall distribution with a Gaussian function, and found to be $(2.7\pm0.2)\%$, which is comparable to the local energy resolution of individual anodes.

\begin{figure}[htbp]
\centering
 \begin{subfigure}[t]{0.495\textwidth}
         \centering
         \includegraphics[width=\textwidth]{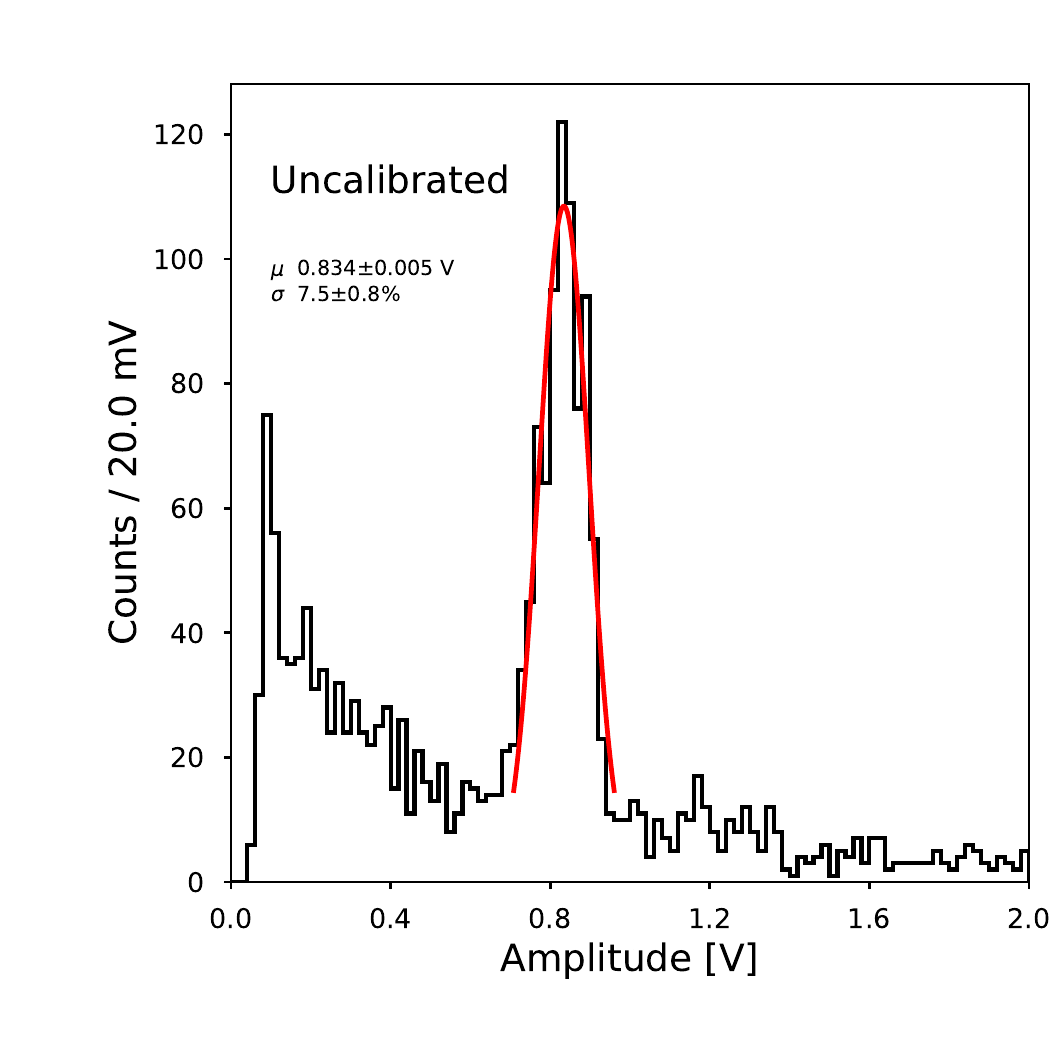}
         \caption{\label{subfig:11chanNoCalSummed}}
     \end{subfigure}
     \begin{subfigure}[t]{0.495\textwidth}
         \centering
         \includegraphics[width=\textwidth]{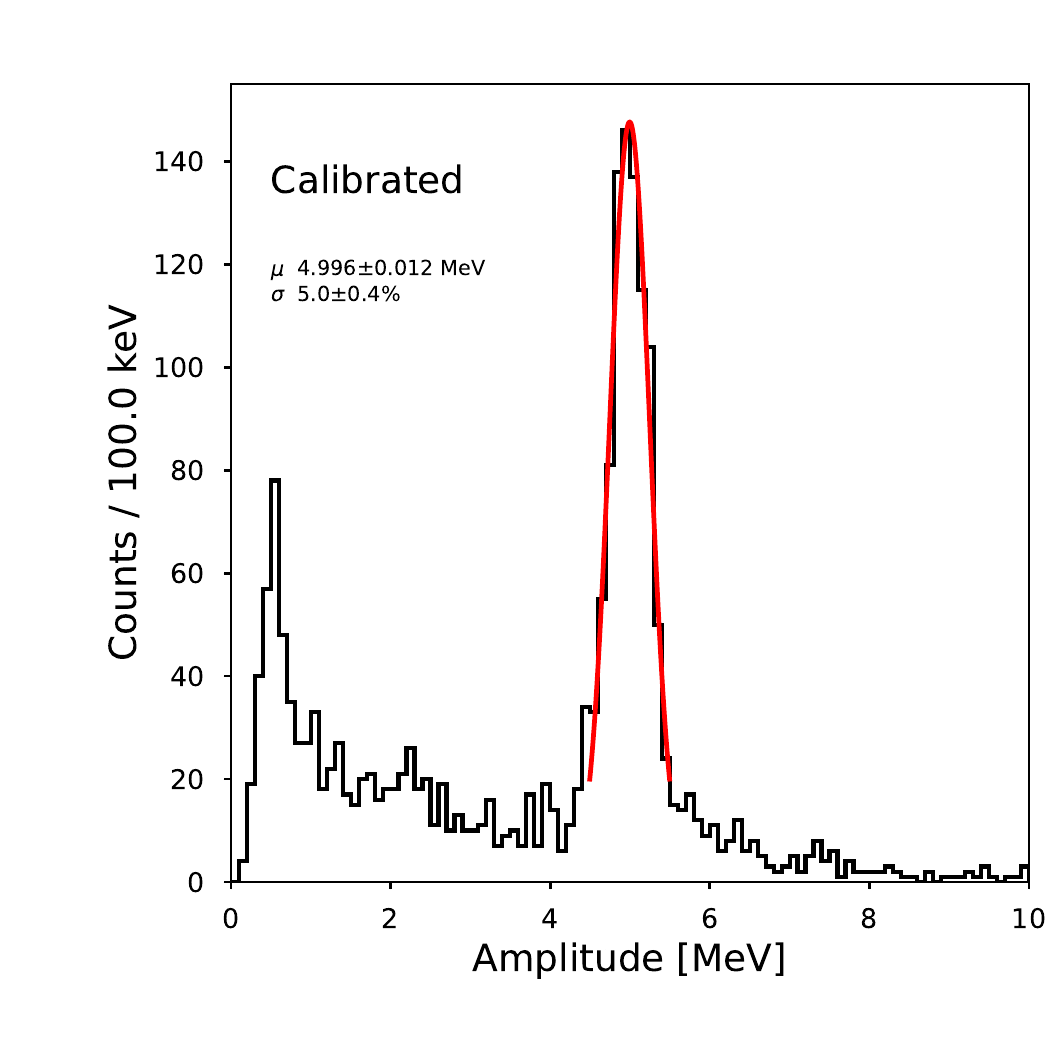}
        \caption{\label{subfig:11chanCalSummed}}
     \end{subfigure}
\caption{\label{fig:midpointTests} Distribution of the sum of recorded amplitudes from each anode in the case where (\subref{subfig:11chanNoCalSummed}) the signals are uncalibrated and (\subref{subfig:11chanCalSummed}) where they are calibrated, when the source was directed between anodes F1 and F5. 
Each distribution is fit with a Gaussian function (red line) to estimate the energy resolution.} 
\end{figure}

Subsequently, the source was directed to a point between two anodes, shown schematically as `P2' in Figure~\ref{fig:achinosCalibrationDiagram}. 
In this case, it was expected that ionisation electrons would more frequently arrive to both anodes. Anodes F1 and F5 were selected as these already exhibited similar gain making an improvement in energy resolution more challenging, thus, testing the individual anode calibration. 
Two cases were considered for analysing this data. In the first, the signal amplitudes from all anodes were summed, being the uncalibrated case, while in the second the calibration derived from Figure~\ref{fig:allAnodescalibComparison} for each anode was first applied before they were summed.
The resulting amplitude distribution for both cases is shown in Figure~\ref{fig:midpointTests}. 
There is a clear improvement in the energy resolution, from $(7.5\pm0.8)\%$ to $(5.1\pm0.4)\%$, when the individual calibration is first applied to the data. However, it can be seen from Figure~\ref{fig:allAnodescalibComparison} that selecting a P2 location between two other anodes would result in an even greater improvement in energy resolution.

\section{Simulation}
\label{sec:simulation}
To study the expected response of ACHINOS, the sensor was modelled using the \texttt{Gmsh}~\cite{gmsh} finite element generator. The mesh is then passed to Elmer~\cite{Ruokolainen_ElmerFEM} finite element solver to compute the electric field in the detector. The field map of the ACHINOS was then provided to a dedicated simulation framework for the spherical proportional counter~\cite{Katsioulas:2019sui}.
The framework incorporates \texttt{Geant4} to compute particle interactions with matter and \texttt{Garfield++} with \texttt{Magboltz} to simulate the drift of charges in gas and calculate the induced signal on the anodes. A further custom simulation module computes the effect of the electronics chain on the signal resulting in a simulated signal that is directly comparable to the experiment.
From this simulation chain, the
gas gain could be computed.

\subsection{Energy resolution}
\begin{figure}[htbp]
\centering
\includegraphics[width=0.3\textwidth]{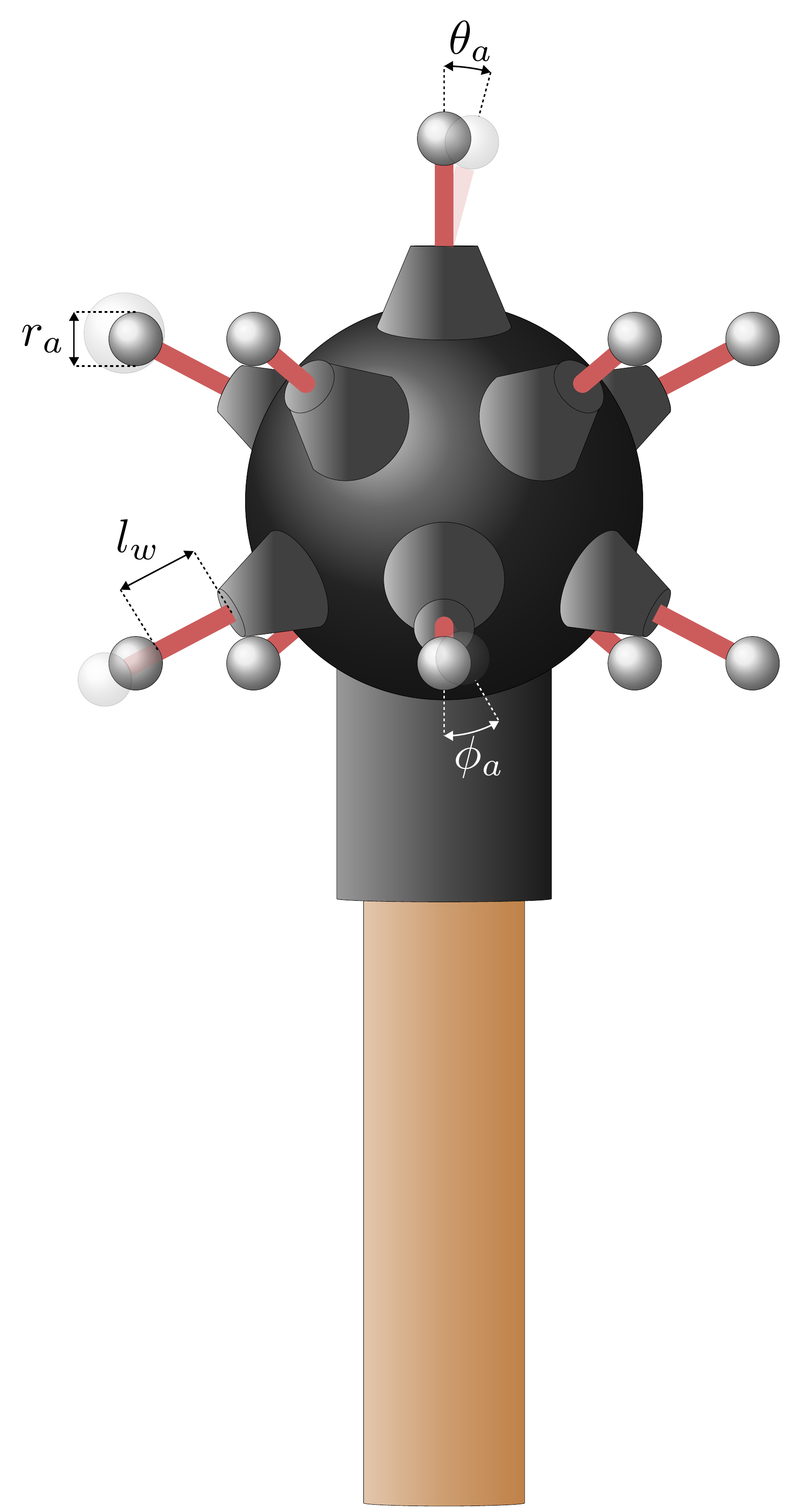}
\caption{\label{fig:achinos_distorted} Schematic of ACHINOS construction variations that were considered in this work. $r_{a}$, $\theta_{a}$, $\phi_{a}$, and $l_{w}$ are the anode radius, anode angle from the nominal position, and anode wire length, respectively.}
\end{figure}
\begin{figure}[htbp]
\centering
\includegraphics[width=0.495\textwidth]{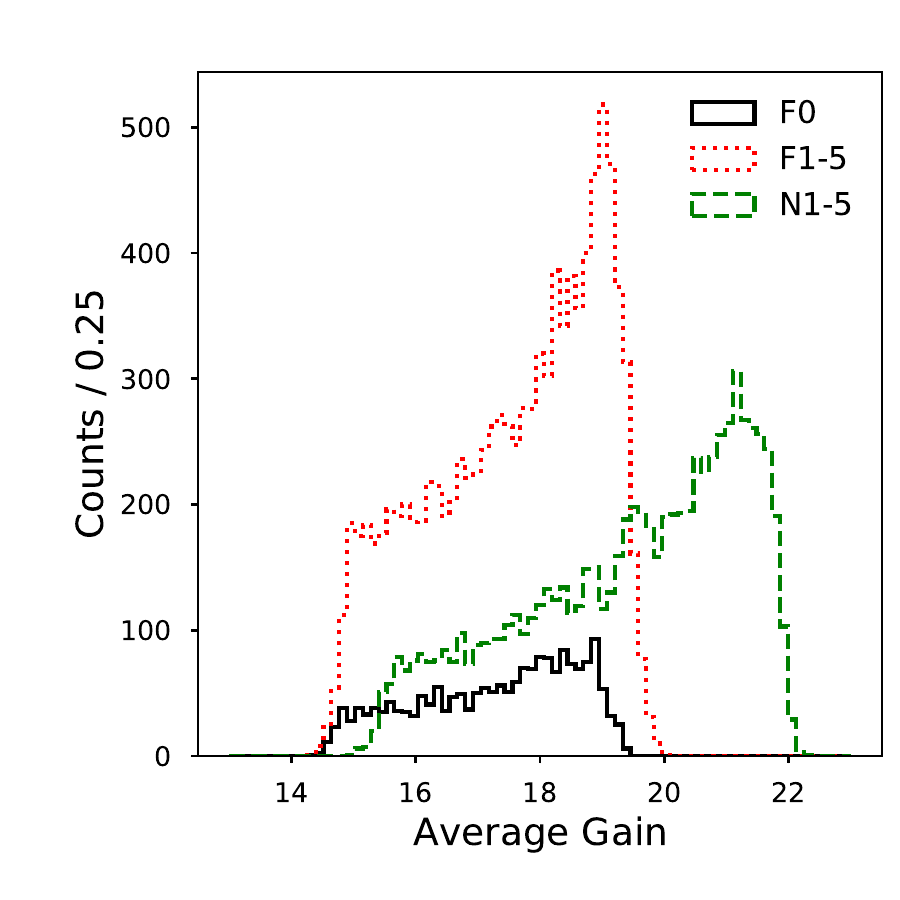}
\caption{\label{fig:gasGain} The gas gain distribution for the F0, F1-5, and N1-5 channels for $l_{w}$ of $2.5\;\si{\milli\meter}$.}
\end{figure}
\begin{figure}[htbp]
\centering
\includegraphics[width=0.495\textwidth]{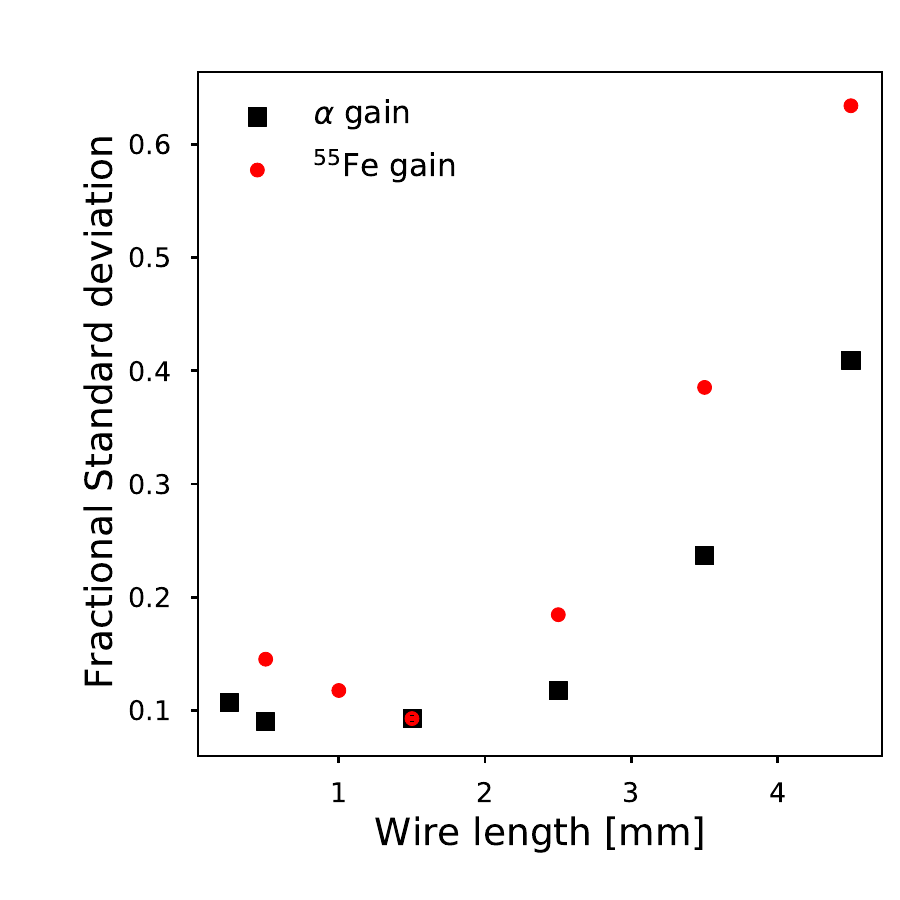}
\caption{\label{fig:wirelength} The standard deviation in the average gain with $l_{w}$ for two voltage configurations; high-gain, used for measuring the energy deposited by x-rays from $^{55}$Fe decays, and lower-gain used when measuring $\alpha$-particles. To ensure the standard deviation in average gain was comparable between each $l_{w}$, the voltage applied was varied to maintain a constant gain. Statistical uncertainties are included but are smaller than the size of the marker.}
\end{figure}

In Ref.~\cite{Katsioulas:2018pyhSUB} the tuning of design parameters for a single-anode spherical proportional counter read-out is discussed, to improve the energy resolution by increasing the electric field homogeneity. In order to set a reference for the ACHINOS energy resolution, a similar simulation study was conducted for the anode wire length $l_{w}$, depicted in Figure~\ref{fig:achinos_distorted}. The study was performed using an 11-anode ACHINOS, with $0.5\;\si{\milli\meter}$ in radius anodes. Electrons were simulated at random initial positions at a fixed radius around the ACHINOS and allowed to drift and diffuse in the detector volume.
For this study, the energy resolution was estimated using the standard deviation in the average gas gain, ignoring avalanche fluctuations and so was sensitive only to gain variations caused by changes in the electric field.
The signal from each anode could be read-out individually in the simulation.  
Figure~\ref{fig:gasGain} shows three example gas gain distributions on the F0, F1-5, and N1-5 channels for $l_{w}$ of $2.5\;\si{\milli\meter}$. The energy resolution was computed at 6 values of $l_{w}$ between $0.25\;\si{\milli\meter}$ and $4.5\;\si{\milli\meter}$ and is plotted against $l_{w}$ in Figure~\ref{fig:wirelength}. A minimum in the energy resolution is found for $l_{w}$ around $1.5\;\si{\milli\meter}$. 

\subsection{Rod-induced gain variation}
\begin{figure}[htbp]
\centering
\includegraphics[width=0.495\textwidth]{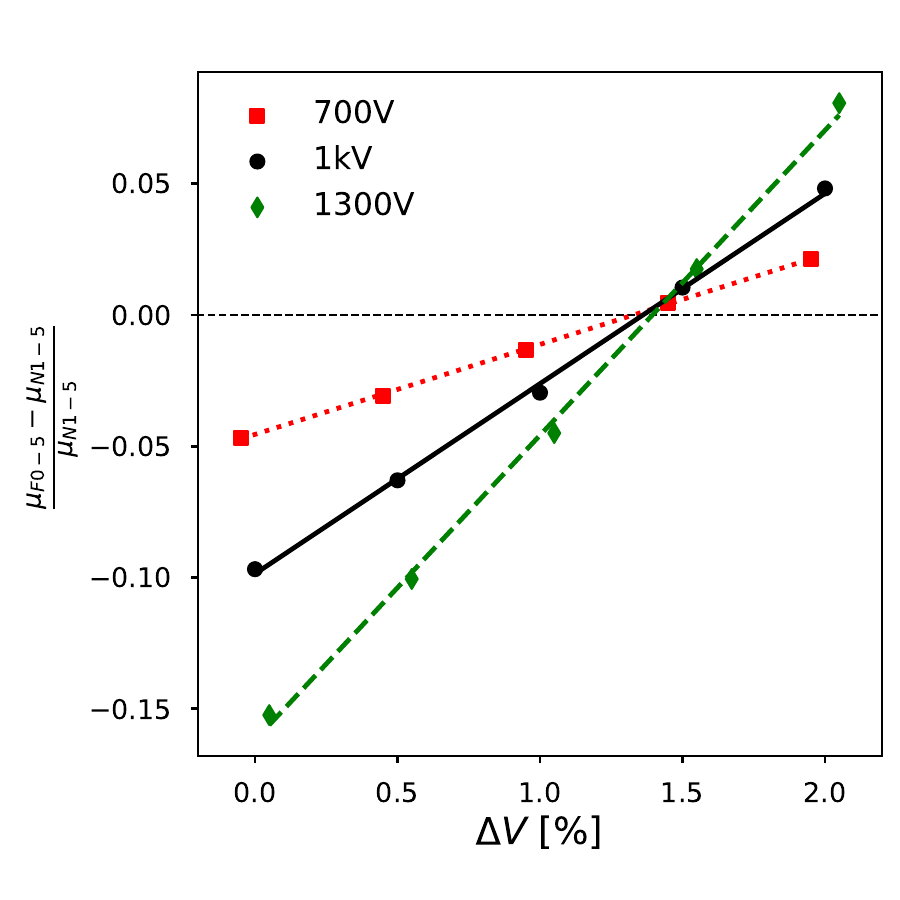}
\caption{\label{fig:near-far-correction}Mean gain difference between the N1-5 and F0-5 anodes ($(\mu_{\text{F0-5}}-\mu_{\text{N1-5}})/\mu_{\text{N1-5}}$) as a function of the additional voltage applied to the Far anodes compared to the Near for three nominal anode voltages. The dashed line is a linear fit to the data which has $(\mu_{\text{F0-5}}-\mu_{\text{N1-5}})/\mu_{\text{N1-5}}=0$ at $\Delta V = (1.38\pm0.02)\%$, $(1.36\pm0.04)\%$, and $(1.34\pm0.05)\%$ for $700$, $1000$, and $1300\;\si{\volt}$, respectively.}
\end{figure}

The geometrical difference between the Near and Far anodes, namely their proximity to the grounded rod, causes an electric field difference and, thus, the dominant gain variation in the nominal ACHINOS. Correcting for this expected gain difference can be achieved by an adjustment in voltage applied to Near and Far anodes. This was studied in the simulation by varying the voltage difference between the N1-5 and F0-5 anodes until the average gain recorded for both was the same. Figure~\ref{fig:near-far-correction} shows the difference in average gain between N1-5 and F0-5 anodes ($(\mu_{\text{F0-5}}-\mu_{\text{N1-5}})/\mu_{\text{N1-5}}$) as a function voltage difference $\Delta V$ between the them for three nominal voltages applied to the anode. The data for each nominal voltage were fit with a linear function, with the best value of $(\mu_{\text{F0-5}}-\mu_{\text{N1-5}})/\mu_{\text{N1-5}}=0$ being comparable, and having a mean of $\Delta V = (1.36\pm0.07)\%$. This was used as the voltage to correct for the nominal Near-Far gain difference in the rest of this section. 

\subsection{Construction imperfections}
In addition to the expected Near versus Far gain difference, construction imperfections of the handmade ACHINOS can cause gain variations of individual anodes. ACHINOS was extensively studied in simulation to understand the effect of construction imperfections on detector operation and how these can be corrected.
Figure~\ref{fig:achinos_distorted} depicts the imperfections that may occur in ACHINOS that were considered in the study: (a) variation in anode radius $r_{a}$; (b) variation in the length of the wire between the anode and central support structure $l_{w}$; (c) displacement of the anode in the polar angle $\theta_{a}$; and (d) displacement of the anode in the azimuthal angle $\phi_{a}$. The manufacturer's tolerance on $r_{a}$ was taken as a representative deviation, while the deviations in the other variables were based on imperfections that could reasonably occur with current construction techniques and materials.  Table~\ref{tab:deformations} summarises the considered deviations. 

\begin{figure}[htbp]
\centering
     \begin{subfigure}[b]{0.495\textwidth}
         \centering
         \includegraphics[width=\textwidth]{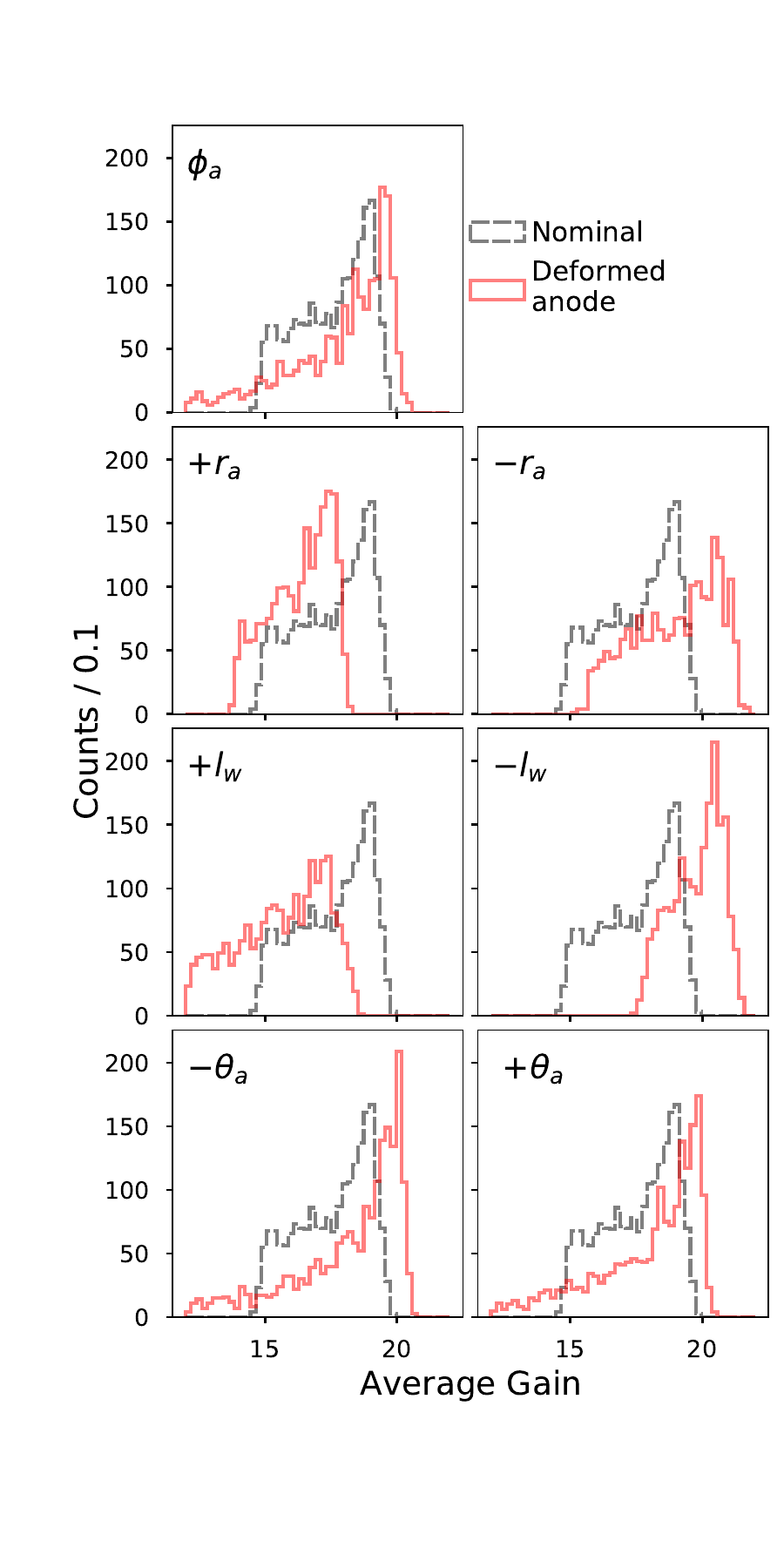}
         \caption{\label{subfig:deformedGain}}
     \end{subfigure}
     \begin{subfigure}[b]{0.495\textwidth}
         \centering
         \includegraphics[width=\textwidth]{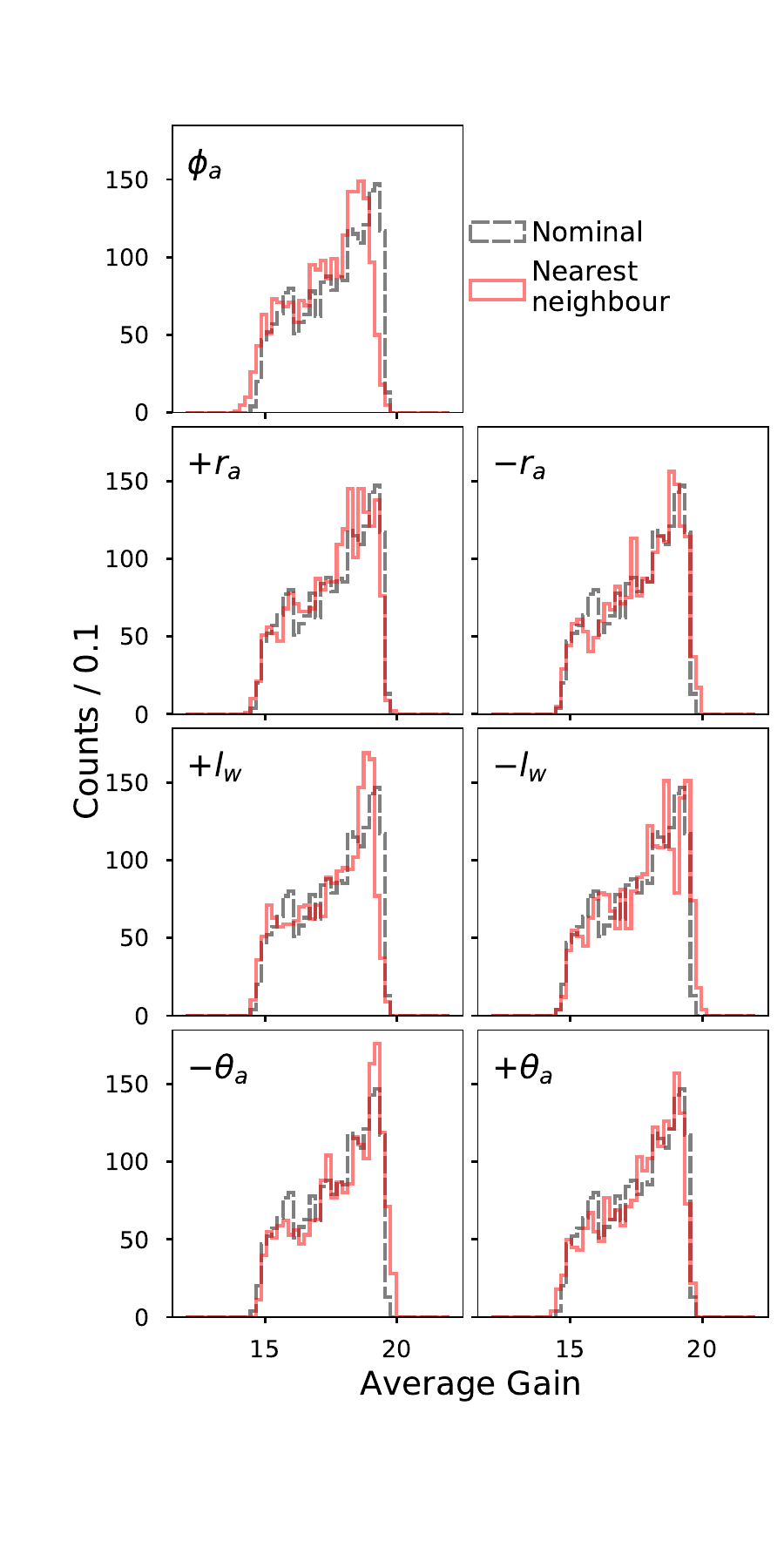}
         \caption{\label{subfig:undeformedGain}}
     \end{subfigure}
\caption{\label{fig:distorted_results} The mean gas gain in each of the deformed cases compared to the nominal, undeformed case, for (\subref{subfig:deformedGain}) the deformed anode and (\subref{subfig:undeformedGain}) its nearest neighbours. The plus ($+$) and minus ($-$) symbols indicate an imperfection is an increase or decrease in that parameter by the tolerance values given in Table~\ref{tab:deformations} respectively.  }
\end{figure}

To form a basis for comparison, a `nominal' ACHINOS was considered with $r_{A}=1.4\;\si{\centi\meter}$, $l_w=2.5\;\si{\milli\meter}$, and $r_a=0.5\;\si{\milli\meter}$. Each of the different imperfections was then simulated on one of the anodes and the results were compared. The effect of each deformation on the mean gain recorded in simulations with electrons placed randomly around the ACHINOS  compared to the nominal are summarised in Figure~\ref{fig:distorted_results}. It can be seen that these deformations can result in an appreciable change in detector gain.

\begin{table}[]
\centering
\caption{\label{tab:deformations}Parameters of the ACHINOS that may vary during its construction that are considered in this study. Parameters are defined in the text and indicated in Figure~\ref{fig:achinos_distorted}.}
\begin{tabular}{llll}
\hline
Parameter&Nominal Value & Deformation Magnitude     \\ \hline
$r_{a}$ & $0.5\;\si{\milli\meter}$          & $\pm0.0125\;\si{\milli\meter}$     \\
$l_{w}$& $2.5\;\si{\milli\meter}$ & $\pm0.05\;\si{\milli\meter}$    \\
 $\theta_{a}$ &    $0\;\si{rad}$  &   $\pm0.5\;\si{rad}$ \\ 
 $\phi_{a}$  &  $0\;\si{rad}$    &    $\pm0.5\;\si{rad}$ \\ \hline
\end{tabular}
\end{table}

\begin{figure}[htbp]
\centering
     \begin{subfigure}[b]{0.495\textwidth}
         \centering
         \includegraphics[width=\textwidth]{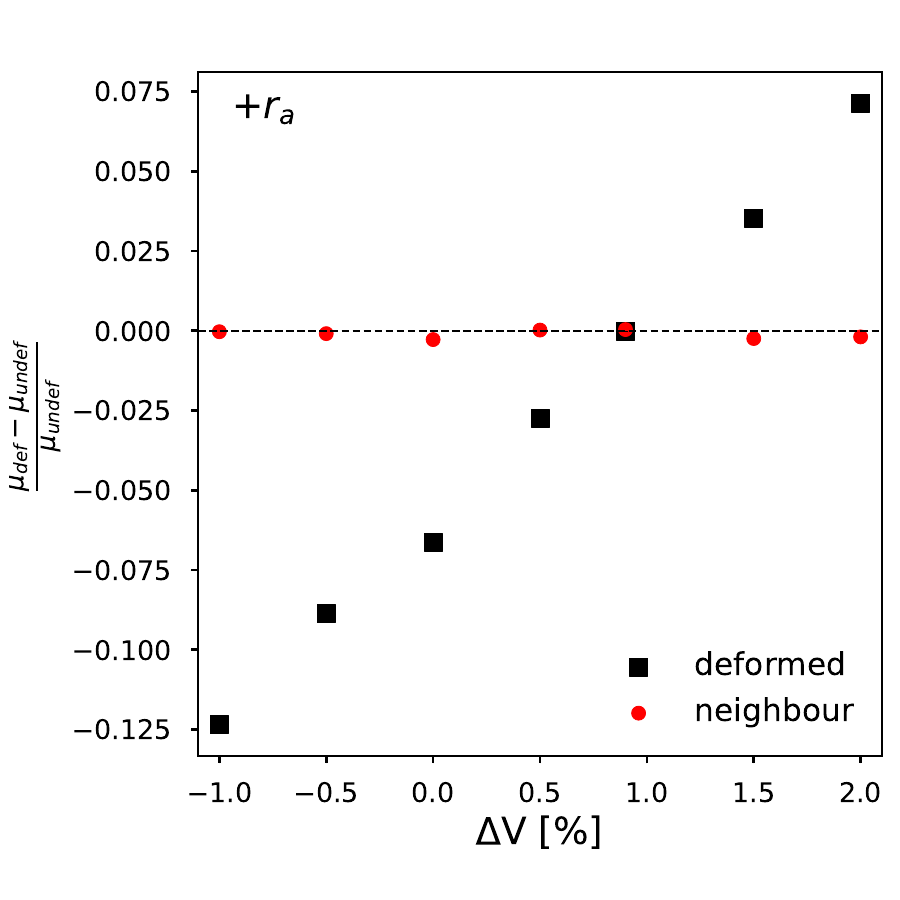}
         \caption{\label{subfig:f1bigstderr}}
     \end{subfigure}
     \begin{subfigure}[b]{0.495\textwidth}
         \centering
         \includegraphics[width=\textwidth]{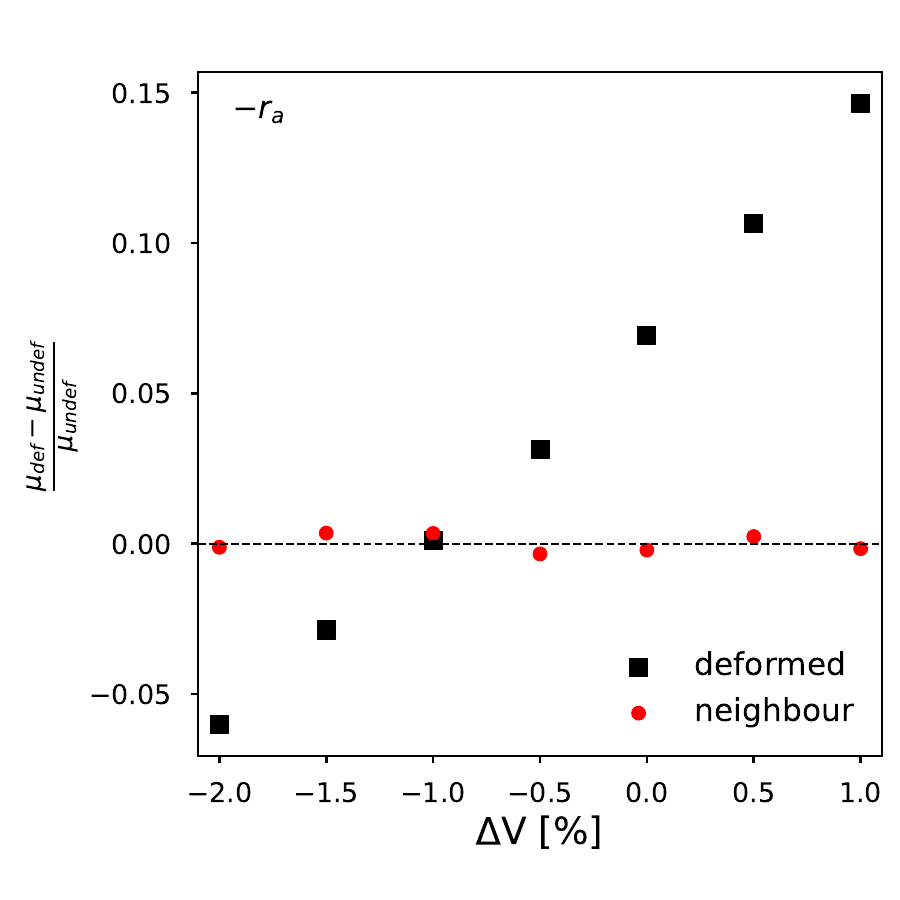}
         \caption{\label{subfig:f1smallstderr}}
     \end{subfigure}
     
     \begin{subfigure}[b]{0.495\textwidth}
         \centering
         \includegraphics[width=\textwidth]{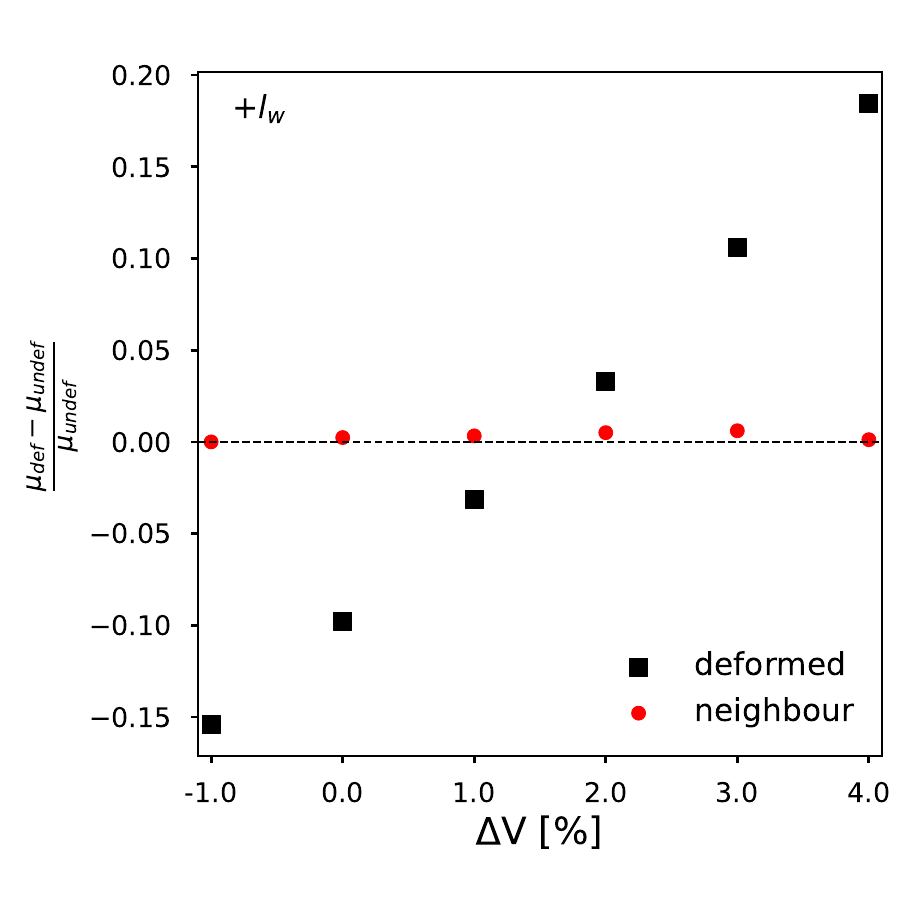}
         \caption{\label{subfig:f1longstderr}}
     \end{subfigure}
     \begin{subfigure}[b]{0.495\textwidth}
         \centering
         \includegraphics[width=\textwidth]{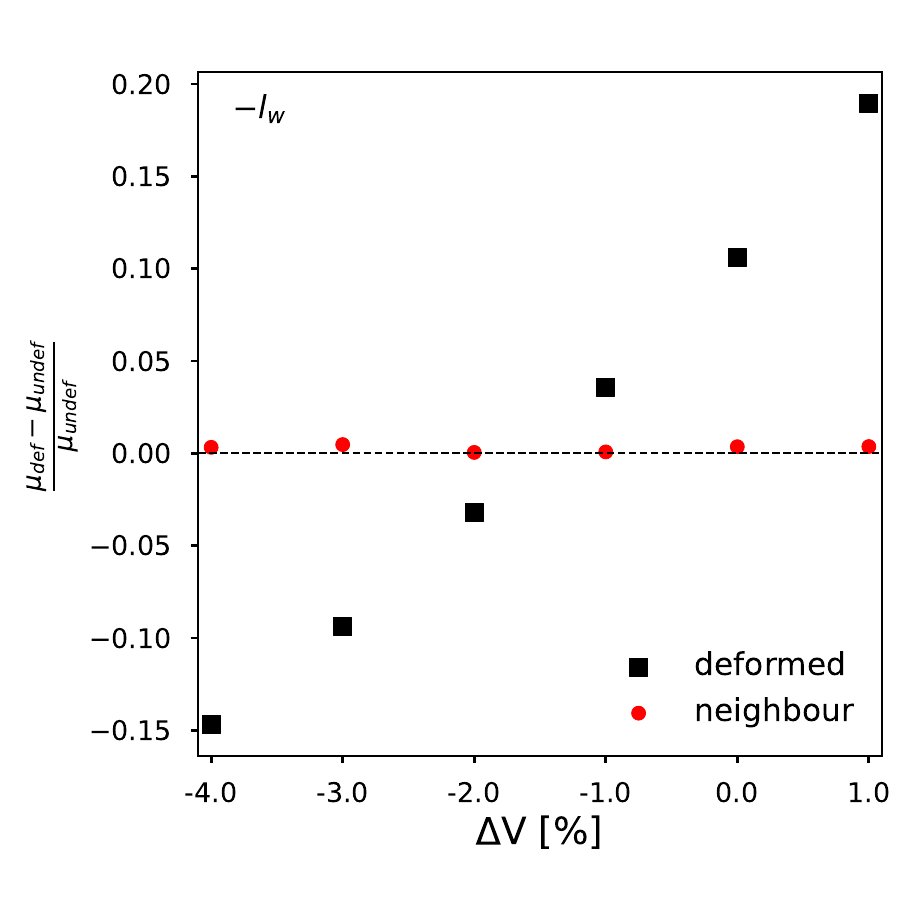}
         \caption{\label{subfig:f1shortstderr}}
     \end{subfigure}
\caption{\label{fig:deformedvoltages} The difference in the mean gain of electrons arriving at the deformed anode and its nearest neighbour, with respect to the nominal gain, as a function of difference in voltage applied to the anode in the case that it is deformed by (\subref{subfig:f1bigstderr}) $r_{a}$ is large,
(\subref{subfig:f1smallstderr}) $r_{a}$ is small,
(\subref{subfig:f1longstderr}) $l_{w}$ is long, and
(\subref{subfig:f1shortstderr}) $l_{w}$ is short. Statistical uncertainties are included but are smaller than the size of the marker.}
\end{figure}

These gain variations can be dealt with using two methods, either calibrating each anode in the analysis stage, or by applying a different voltage to each anode during data-taking. The required adjustment of the voltage to compensate for each deformation was studied using the simulation, with Figure~\ref{fig:deformedvoltages} showing the deviation in the mean gain distribution as a function of difference in voltage applied to the deformed anode for each case $(\mu_{\text{def}}-\mu_{\text{undef}})/\mu_{\text{undef}}=0$. The voltage difference at which $(\mu_{\text{def}}-\mu_{\text{undef}})/\mu_{\text{undef}}=0$ indicates the voltage difference at which the deformation is corrected for. The gain of a neighbouring anode to the deformed one is shown for each of the voltage scenarios, and shows little variation from the nominal, demonstrating that the voltage adjustment does not affect the other anodes. 

\section{Summary}
Individual anode read-out and calibration for the multi-anode ACHINOS sensor has been developed. First results with an individually read-out ACHINOS show significant improvement in energy resolution with individual-anode calibration compared to two-channel read-out. This development is important for several applications of the spherical proportional counter, for example, by enabling improved background rejection for direct dark matter searches through track and position reconstruction and improved detector fiducialisation.
The sources of anode-by-anode gain variations have been investigated, and extensive simulations of the ACHINOS have been used to study how the geometry and construction imperfections affect the energy resolution of both the deformed anode and a neighbouring anode. The voltage required to correct for gain differences caused by geometry or anode deformations was also studied, and found to be of order $1\%$ for all considered scenarios.

\acknowledgments

This project has received funding from the European Union’s Horizon 2020 research and innovation programme under the Marie Skłodowska-Curie grant agreements No 845168 (neutronSphere), No 841261 (DarkSphere) and No 101026519 (GaGARin). Support from the UK Research
and Innovation - Science and Technology Facilities
Council (UKRI-STFC) is acknowledged; KN is supported by the University of Birmingham Particle
Physics consolidated grant No. ST/W000652/1; PK is supported by  ST/X005976/1; LM is supported by ST/X508913/1. KN acknowledges support by the Deutsche Forschungsgemeinschaft (DFG, German Research Foundation) under Germany’s Excellence Strategy – EXC 2121 „Quantum Universe“ – 390833306.

\clearpage

\bibliographystyle{JHEP}
\bibliography{mybib.bib}

\end{document}